\documentclass[aps,prx,superscriptaddress,showpacs,preprint]{revtex4-2}
\usepackage{graphicx}
\usepackage{amsmath}
\usepackage{amssymb}
\usepackage[utf8]{inputenc}
\usepackage{comment}
\usepackage{caption}
\usepackage{subcaption}
\usepackage{dcolumn}
\usepackage{bm}
\usepackage{hyperref}
\usepackage{color}
\usepackage{soul}
\newcommand{\ha}{\hat{a}}

\newcommand{\had}{\hat{a}^{\dagger}}
\newcommand{\hH}{\hat{H}}

\newcommand{\hn}{\hat{n}}

\newcommand{\bG}{\mathbf{G}}

\newcommand{\co}{\textrm{c}}

\newcommand{\be}{\begin{equation}}
\newcommand{\ee}{\end{equation}}

\newcommand{\Tr}{\mathrm{Tr}}

\newcommand{\Rmnum}[1]{\expandafter\@slowromancap\romannumeral  #1@}

\newcommand{\nh}{\hat{n}}

\newcommand{\KL}{\mathrm{K}}
\newcommand{\LW}{\mathrm{LW}}
\newcommand{\HF}{\mathrm{HF}}
\newcommand{\SB}{\mathrm{2B}}
\newcommand{\GW}{\mathrm{GW}}
\newcommand{\Tm}{\mathrm{T}}
\newcommand{\mx}{\mathrm{x}}
\newcommand{\hN}{\hat{N}}
\newcommand{\tG}{\tilde{G}}

\UseRawInputEncoding

\begin{document}

\title{Potential energy surfaces from many-body functionals: 
analytical benchmarks and conserving many-body approximations
}


\author{Giovanna Lani} 
\affiliation{Theory and Simulation of Materials (THEOS), \'Ecole Polytechnique F\'ed\'erale de Lausanne, CH-1015 Lausanne, Switzerland}
\author{Nicola Marzari} 
\affiliation{Theory and Simulation of Materials (THEOS), \'Ecole Polytechnique F\'ed\'erale de Lausanne, CH-1015 Lausanne, Switzerland}
\affiliation{National Centre for Computational Design and Discovery of Novel Materials (MARVEL), \'Ecole Polytechnique F\'ed\'erale de Lausanne, CH-1015 Lausanne, Switzerland}
\affiliation{Laboratory for Materials Simulations (LMS), Paul Scherrer Institut, CH-5232 Villigen, Switzerland}
\date{\today}


\begin{abstract}
We investigate analytically the performance of many-body energy functionals, derived respectively by Klein and Luttinger and Ward, at different levels of diagrammatic approximations, ranging from second Born, to GW, to the so-called T-matrix, for the calculation of total energies and potential energy surfaces.
We benchmark our theoretical results on the extended two-site Hubbard model, which is analytically solvable and for which several exact properties can be calculated. Despite its simplicity, this model displays the physics of strongly correlated electrons: it is prototypical of the H$_2$ dissociation, 
a notoriously difficult problem to solve accurately for the majority of mean-field based approaches. 
We show that both functionals exhibit good to excellent variational properties, particularly in the case of the Luttinger-Ward one, which is in close agreement with fully self-consistent calculations, and elucidate the relation between the accuracy of the results and the different input one-body Green's functions.
Provided that these are wisely chosen, we show how the Luttinger-Ward functional can be used as a computationally inexpensive alternative to fully self-consistent many-body calculations, without sacrificing the precision of the results obtained.
Furthermore, in virtue of this accuracy, we argue that this functional can also be used to rank different many-body approximations at different regimes of electronic correlation, once again bypassing the need for self-consistency.  
\end{abstract}

\maketitle

\section{Introduction}
Many-body energy functionals, such as the one proposed by Luttinger and Ward (LW)~\cite{LW-1960}, or the one by Klein~\cite{Kle-PR-61}, or the more recent $\Psi$ functional, originally derived in the work of Almbladh \textit{et al.} \cite{Alm-Int-99,Hindegren-thesis}, have been employed starting from the 90's for total energy calculations in a limited number of systems, ranging from the homogeneous electron gas (HEG) in \cite{Hindegren-thesis} and in \cite{You-JPA-03}, to atoms and small diatomic molecules \cite{Dah-JCP-04,Dah-PRB-04,Dah-PRA-06}.
%
These first results showed great promise, since the functionals used at various level of many-body approximations, yielded comparable accuracy to state-of-the-art quantum Monte Carlo (QMC) calculations, but at much cheaper computational costs. The strength of these functionals lies in their variational properties, which allow to bypass the need of an expensive and sometimes technically difficult fully self-consistent many-body calculation in order to obtain total energies (and other properties as well).
An investigation was performed for example in \cite{Dah-PRA-06} for a small set of spherically symmetric atoms:
employing the so-called GW approximation \cite{Hed-PR-65}, the authors obtained results comparable in accuracy to those obtained by solving self consistently the Dyson equation, but at a much reduced cost, since their procedure amounted to a one-step post-processing calculation. 
Furthermore, more fundamental insights were obtained; for instance, the superior variational character of the LW functional over the Klein one, 
or the effect of using more correlated Green's functions (e.g., from Hartree-Fock (HF) rather than Kohn-Sham local density approximation (LDA)).
The conclusions were later supported by the work of You \textit{et al.} \cite{You-JPA-03} on the homogeneous electron gas (HEG), where the authors obtained total energies of QMC quality using HF Green's functions as input for the functionals. 
%
A handful of more recent studies \cite{Ary-PRL-02,Miy-PRB-02,Car-PRL-13} have also benchmarked the Klein functional. In these works, total energies have been calculated at the random-phase approximation (RPA) level using KS-LDA Green's functions, and they were found to be in reasonable agreement with high level approaches, but the question on whether alternative input Green's functions could improve on the results was not addressed. \\
Lately, there has been a renewed interest in the use of many-body functionals, since they are by no means restricted to the evaluation of zero temperature properties. 
In fact, if one is interested in including finite-temperature effects in electronic structure calculations, the most natural way is precisely to resort to them, because of their straightforward connection to the grand canonical potential \cite{SteLeeu}.
In \cite{Wel-JCP-16}, the authors employ the LW functional (in its finite-temperature formulation) to evaluate self-consistently thermodynamical quantities for a 1D hydrogen chain. 
In a related work \cite{Pok-JCP-21}, many-body functionals were used indirectly and in an original way to obtain two-particle properties, such as $\langle S^2 \rangle$ or $\langle N^2 \rangle$, bypassing the expensive solution of the Bethe-Salpeter equation.
Furthermore, within the dynamical mean-field theory community~\cite{DMFT-rev}, the LW functional has long been discussed, and it has also started to be employed in calculations on correlated materials. 
For instance, in \cite{Hau-PRB-16}, the free energy functional is differentiated with respect to the atomic displacements, in order to carry out structural optimizations. The formalism was also later extended to the use of ultra-soft pseudo-potentials~\cite{Ple-PRB-21}, in order to broaden its applicability to plane-wave based calculations.
In a similar spirit, in another recent work \cite{chiarotti2023energies}, a newly derived many-body functional was employed to compute the equation of state of SrVO$_3$.
Also within the reduced-density-matrix community there have been ongoing research efforts. Recently, in the work of Giesbertz and coauthors \cite{Gie-EPJB-18}, the functionals where first generalised to be used in the space of one-body reduced density matrices and then applied to the computation of the bonding curve of a model hydrogen molecule.
In parallel to the aforementioned works, there have been intense research efforts geared towards benchmarking Green's function methods either on model or very simple systems, in order to gain some fundamental and conceptual insights.  
For example, in~\cite{Riva-PRL-23}, the authors calculated the photo-emission spectra coupling the one-body Green's function with the three-body one, and applied their approach to the Hubbard dimer.
In~\cite{criso24} Crisostomo and coworkers put forward an original approach to extract the exchange-correlation energy of DFT from the Galitskii-Migdal formula, which is based on many-body Green's functions, and test it both on the Hubbard dimer and the HEG. 
Finally, in~\cite{loos24}, the authors investigated the performance of the GW approximation for the challenging case of very simple multireference systems, where strong electronic correlation plays a crucial role. 
Despite the early works and the more recent ones, this research topic still presents many open questions and there is a large uncharted territory both at the fundamental and at the application level. 
Within the former, only two members of the infinite class of many-body functionals have been explored and only for a couple of simple many-body approximations, that is the second Born and the GW one. Within the latter, apart from a few notable exceptions \cite{Ary-PRL-02,chiarotti2023energies}, there has never been an application to solids, which would instead be of great interest. \\
 \noindent This work aim is to gain a fundamental understanding of the most well-known functionals when employed in conjunction with different combinations of their many-body ingredients. The analytical nature of the study will provide exact benchmarks and will complement and clarify earlier numerical works that preceded it. 
\section{General theory of many-body functionals}
 \label{sec:sec2}
 \subsection{The Klein and Luttinger Ward functionals}
Before going into the main results of this paper, we shall introduce the Klein \cite{Kle-PR-61,Baym-1962} and the LW functionals \cite{LW-1960} and their many-body ingredients. Following \cite{Dah-JCP-04,Dah-PRB-04,Dah-PRA-06}, we use Matsubara Green's functions evaluated in the zero-temperature limit, in which the Matsubara sums can be replaced by frequency integrals. This has the advantage of being adapted to total energy calculations, while at the same time reducing the necessary algebra as compared to the zero-temperature formalism.

At finite temperature the Klein functional for the grand potential $\Omega$ is defined as:
\be
\Omega_\KL [G] = - \Tr \ln( - G^{-1} ) - \Tr (G_0^{-1} G -1) + \Phi [G], 
\label{eq:Klein}
\ee
where we used the definition of the trace as in \cite{Dah-PRA-06}, namely $\Tr = \sum_{\sigma} \int \frac{d\omega}{2\pi} \sum_i$ where $\sigma$ is a spin index and $i$ a space one. The minus sign in front of $G^{-1}$ in the logarithm ensures that 
with the standard choice of logarithmic branch cut the trace is well-defined \cite{LW-1960,Kar-PRB-16}.
$\Phi [G]$ is a functional of the one-body Green's function \cite{SteLeeu} with the property that $\delta \Phi/ \delta G = \Sigma [G]$ which is the many-body self-energy. In practical applications this functional needs to be approximated via a selection of Feynman diagrams of vacuum type.
In the zero-temperature limit, it can be related to the total energy in the following fashion:
\be
E_\KL [G] = \mu N + \Omega_\KL [G], 
\ee
where $\mu$ is the chemical potential, $N$ the particle number and $\Omega_\KL$ the grand potential of Eq.~\ref{eq:Klein}.
When making a variation $\delta G$ in the functional, its first order change is given by \cite{Baym-1962,Dah-PRA-06}:
\be
\delta \Omega_\KL = \Tr ( G^{-1} - G_0^{-1} + \Sigma [G] ) \delta G
\ee
and the first order derivative vanishes whenever:
\be
G=G_0 + G_0 \Sigma[G] G \,\,\,\Leftrightarrow \,\,\, G^{-1} = G_0^{-1} - \Sigma [G],
\ee
which holds when $G$ solves the Dyson equations self-consistently with a so-called $\Phi$-derivable self-energy $\Sigma [G]$ \cite{Baym-1962}.
Using the Dyson equation in the first term of Eq.~\ref{eq:Klein} yields a new form for $\Phi$, known as the Luttinger-Ward functional, defined as \cite{Dah-PRA-06}: 
\be
\Omega_\LW [G] = - \Tr \ln ( \Sigma [G] - G_0^{-1} ) - \Tr ( G \Sigma [G]) + \Phi [G].
\label{eq:LW}
\ee
Whenever the input $G$ inserted in the functionals satisfies the Dyson equation self-consistently, $\Omega_\KL [G] =\Omega_\LW [G]$ 
by construction. 
However, for any other approximate input Green's function $\tG$, $\Omega_\KL [\tG] \neq \Omega_\LW [\tG]$.
It can be readily checked that also $\Omega_\LW [G]$ is stationary for the $G$ that solves the Dyson equation self-consistently \cite{Dah-PRA-06}. 
It has also been argued \cite{Dah-PRB-04,Dah-PRA-06} that since the LW functional is obtained by using the Dyson equation once iteratively, it should be more accurate than the Klein one, given the same input Green's function.
Continuing by induction, it is easy to see that there is an infinite number of variational functionals, whose different expressions can be constructed following, e.g., the prescription given in \cite{Bar-PRB-05} and also in \cite{SteLeeu}. 
In this work we will restrict ourselves to the study of the Klein and LW ones, which have already proven to be useful in some practical applications \cite{Dah-PRA-06,Ary-PRL-02,Wel-JCP-16, Pok-JCP-21}.
Both functionals exhibit a number of important properties, which we will illustrate below. \\
First, whenever $G$ is a self-consistent solution of the Dyson equation with a $\Phi$-derivable self-energy, and 
$\tG$ an approximate input Green's function, with $\Delta G = \tG-G$, we have that:
\be
\Omega [ \tG] = \Omega [G + \Delta G] = \Omega [G] + O ((\Delta G)^2),
\label{DeltaG_exp}
\ee
since the first derivative vanishes at the self-consistent $G$. This means that if $\Delta G$ is small enough, we only make an error of
second order in the energy when using an approximate input $\tG$, for example obtained from a Kohn-Sham or a Hartree-Fock calculation.
This opens up the possibility of calculating
total energies with comparable accuracy to those calculated self-consistently but using a substantially simpler procedure, thereby saving significant computational effort.\\
Second, as already mentioned, the Klein and LW forms of the functionals are equal when evaluated from a self-consistent solution of the Dyson equation.
This implies that if for an approximate $\tG$, $\Omega_\KL [\tG]$ is not close to $\Omega_\LW [\tG]$, one can conclude
that the input Green's function is far from the self-consistent one, and the converse is also true.
This feature can be used as a test of {\em closeness to self-consistency}.
This aspect will be discussed later in this manuscript for the benchmark system of the Hubbard dimer.
Due to the aforementioned properties, both functionals show great promise as a cheap and at the same time accurate tool for total energies calculations. \\
In the remaining of this paper, we will present a {\em fully analytical} study of both the Klein and the LW functionals for such an exactly solvable model, which will provide us with a clear understanding of their performance over a wide regime of electronic correlation.

                    
\subsection{Relation between the functionals for an arbitrary input Green's function}

Let us first consider an arbitrary input Green's function $\tG$. The LW functional can be recast as:
\be
\Omega_\LW [\tG] = - \Tr \ln ( \Sigma [\tG] - G_0^{-1} ) - \Tr ( \tG \Sigma [\tG]) + \Phi [\tG].
\label{eq:LW2}
\ee
It is convenient to split the self-energy into a Hartree-Fock (HF) and a correlation part:
\be
\Sigma [G] = \Sigma_{\HF} [G] + \Sigma_\co [G],
\ee
and to further define a Green's function $\bG$ which represents the first iteration of the Dyson equation towards the HF Green's function \cite{Dah-JCP-04,Dah-PRA-06}, starting
from a self-energy evaluated at the input $\tG$, or more precisely:
\be
\bG = G_0 + G_0 \Sigma_\HF [\tG] \bG \,\, \leftrightarrow \,\, \, \bG^{-1} = G_0^{-1} - \Sigma_\HF [\tG].
\label{eq:Gbar}
\ee
Then, for the logarithmic expression in the first term of Eq.(\ref{eq:LW2}), one can write:
\begin{align}
\Tr \ln ( \Sigma [\tG] - G_0^{-1} ) &= 
\Tr \ln ( \Sigma_\HF [\tG]  + \Sigma_\co [\tG] - G_0^{-1} )  \\
&= \Tr \ln (- \bG^{-1} ) + \Tr \ln (1- \bG \Sigma_\co [\tG]),
\end{align}
while for the second term: 
\begin{align}
\Tr ( \tG \Sigma [\tG])  &= \Tr (\tG \Sigma_\HF [\tG]) +   \Tr (\tG \Sigma_\co [\tG]) \nonumber \\
&= \Tr (G_0^{-1} \tG - 1) + \Tr (1- \bG^{-1} \tG ) \nonumber \\
&+   \Tr (\tG \Sigma_\co [\tG]). 
\end{align}
Using the above equations, one can express the LW functional as:

\be
\Omega_\LW [ \tG ] = \Omega_\KL [\tG] + C_\LW [\tG], 
\label{eq:LW_correction1}
\ee
where the LW correction term has been defined as:
\begin{align}
C_\LW [\tG] =  &- \Tr (\tG \Sigma_\co [\tG])  - \Tr \ln (1-\bG \Sigma_\co [\tG]) \nonumber \\
&+ \Tr \ln (-\tG^{-1}) - \Tr \ln (-\bG^{-1}) \nonumber \\
&- \Tr (1- \bG^{-1} \tG ).
\label{eq:LW_correction2}
\end{align}

Eqs.(\ref{eq:LW_correction1}) and (\ref{eq:LW_correction2}) are the main result of this section and play a key role for evaluating the LW functional written in terms of the Klein one plus a correction term. 
In the special case of $\tG=G_\HF$, from Eq.(\ref{eq:Gbar}) it follows that $\bG=\tG=G_\HF$, which causes the second line of Eq.(\ref{eq:LW_correction2}) to vanish. This implies that:

\begin{align}
\Omega_\LW [ G_\HF ] &= \Omega_\KL [G_\HF] + C_\LW [ G_\HF] \\ 
C_\LW [ G_\HF] &= - \Tr (G_\HF \Sigma_\co [G_\HF])  \nonumber \\
&- \Tr \ln (1- G_\HF \Sigma_\co [G_\HF]),
\label{Eq:LW-corr}
\end{align}
Expanding 
the logarithmic term of Eq.\ref{Eq:LW-corr}, one has:
\be
C_\LW [ G_\HF]  = \frac{1}{2} \Tr ( G_\HF \Sigma_\co [G_\HF] G_\HF \Sigma_\co [G_\HF]) + O (\Sigma_c^3),
\ee
which is of second order in the correlation self-energy. 
This shows that whenever $G_\HF$ is a good guess to the self-consistent $G$, $\Sigma_\co$ will be small, and so will the correction $C_\LW $, and as a consequence the Klein and LW functionals will yield results very close to each another. 

                  
\subsection{A general input $\tG$}

In this section we consider a spin compensated system with an arbitrary number of particles, and construct an input Green's function that is diagonal in the HF basis, but still differs from it due to a shift in its poles; this allows to take into account in a very simple way some correlation effects. The HF Green's function and the model one thus read:

\begin{align}
G_{\HF,kl} ( i \omega) &= \frac{\delta_{kl}}{i \omega- \epsilon_{\HF,k} + \mu } \\
\tilde{G}_{kl} ( i \omega) &= \frac{\delta_{kl}}{i \omega- \epsilon_{k} + \mu } 
\end{align}
where the values $\epsilon_k$ need to be suitably chosen (an explicit example is given later for the Hubbard dimer). 
Here we consider a closed-shell system with $N=2M$ particles and which each level with orbital energy
$\epsilon_i$ is either doubly occupied or zero (some of the $\epsilon_i$ may be degenerate). 
The fully spin-dependent Green's function is then diagonal in spin $\tG_{i\sigma,j\sigma'}=\delta_{\sigma \sigma'} \tG_{ij}$.
The chemical potential is chosen in such a way that the system has the correct number of particles $N$, i.e.,
\be
N =-i \,  \Tr (\tilde{G}) = 2 \sum_{j=1}^\infty \theta (\mu-\epsilon_j) 
\ee
where $\theta (x)$ is the Heaviside function, i.e., $\theta (x)=1$ if $x>0$ and zero otherwise.
We further assume that the chemical potential $\mu$ can be chosen such that both $G_\HF$ and $\tG$ have the same particle number $N=2M$, where the lowest $M$ levels
are occupied for both Green's functions. So $\mu$ becomes:
\be
\epsilon_{M} < \mu < \epsilon_{M+1} \quad \quad \epsilon_{\HF,M} < \mu < \epsilon_{\HF,M+1} 
\ee
which is always possible whenever
\be
\max{( \epsilon_{M} , \epsilon_{\HF,M})} < \min{ ( \epsilon_{M+1} , \epsilon_{\HF,M+1})}
\ee
Now, since $\tG$ is diagonal in the HF basis with the same particle number, it follows that the one-particle density matrix calculated from $\tG$ and $G_\HF$ is the same
such that 
\be
\Sigma_\HF [\tG ] = \Sigma_\HF [ G_\HF] \quad \Rightarrow \quad\bG = G_\HF. 
\ee
Armed with the above expressions, one can recast the many-body functionals in a more convenient way. 
The Klein functional now reads:
\begin{align}
\Omega_\KL [\tG ] = - \Tr \ln ( - \tG^{-1}) - \Tr ( G_0^{-1} \tG-1 ) + \Phi [\tG].
\end{align}
More specifically, the first term becomes \cite{SteLeeu} : 
\begin{align}
- \Tr \ln ( - \tG^{-1})&= \sum_j 2(\epsilon_j - \mu) \theta (\mu-\epsilon_j)  \nonumber \\
&= -\mu N + \sum_j 2 \epsilon_j \, \theta (\mu-\epsilon_j),  
\end{align}
while the second:
\begin{align}
- \Tr  (G_0^{-1} \tG-1 ) &= 
- \Tr ( \Sigma_\HF [G_\HF] \tG) - \Tr  (G_\HF^{-1} \tG-1 ) \nonumber \\
&= 2 \sum_j (\epsilon_{\HF,j} - \Sigma_{\HF,jj} - \epsilon_j  ) \theta (\mu-\epsilon_j);
\end{align}
collecting all the terms one arrives at:
\be
 \Omega_\KL [\tG ] = -\mu N + 2 \sum_j (\epsilon_{\HF,j} - \Sigma_{\HF,jj}   ) \theta (\mu-\epsilon_j) + \Phi [\tG].
\ee
Equivalently, since the total energy is given by $E_\KL = \mu N + \Omega_\KL$, it holds that:
\be
E_\KL [\tG] =  2 \sum_j (\epsilon_{\HF,j} - \Sigma_{\HF,jj}   ) \theta (\mu-\epsilon_j) + \Phi [\tG]
\label{eq:energy_Klein}
\ee
It now only remains to calculate the LW correction of Eq.(\ref{eq:LW_correction2}). 
Using that $\bG=G_\HF$, one has that:
\begin{align}
\displaystyle
\Tr \ln (-\tG^{-1}) &- \Tr \ln (-\bG^{-1}) - \Tr (1- \bG^{-1} \tG ) \nonumber \\
=& 2 \sum_j ( \epsilon_{\HF,j} - \epsilon_j) \theta (\mu-\epsilon_j)  \nonumber \\
&+ 2 \sum_j (\epsilon_j - \epsilon_{\HF,j} ) \theta (\mu-\epsilon_j)  = 0,
\end{align}
and the correction for our input Green's function becomes:
\be
C_\LW [\tG] =  - \Tr (\tG \Sigma_\co [\tG])  - \Tr \ln (1-G_\HF \Sigma_\co [\tG]) 
\ee
In summary, for the model input Green's function $\tG$, the following three expressions for the functionals are valid:

\begin{align}
E_\KL [ \tG ] &= 2 \sum_j (\epsilon_{\HF,j} - \Sigma_{\HF,jj}   ) \theta (\mu-\epsilon_j) + \Phi [\tG] 
\label{eq:energy_Klein} \\
E_\LW [\tG] &= E_\KL [ \tG ] + C_\LW [\tG] \\
C_\LW [\tG] &=  - \Tr (\tG \Sigma_\co [\tG])  - \Tr \ln (1-G_\HF \Sigma_\co [\tG]) 
\label{eq:LW_correction3}
\end{align}

These are the key equations that will be used in the remaining of the paper.

                  
\section{Application to the extended Hubbard dimer}

\subsection{The Hamiltonian}
\subsubsection{Extended Hubbard dimer and H$_2$ model}
\label{extended_dimer}

As a first application of the general theory presented in the previous section, we study the so-called extended Hubbard dimer. It was demonstrated in \cite{Gie-EPJB-18} that the parameters in the model can be chosen in such way that they provide a very good representation of the
H$_2$-molecule bonding curve. The details of this are given in Appendix \ref{Hubbard_parameters}.
The model allows for a {\em fully analytic} study of the variational functionals in the scenario of a homolitic molecular dissociation and provides a detailed insight of the parameter dependencies, that would be very challenging to gain from a
purely numerical study. \\

The Hamiltonian of the dimer has the following explicit form:
\begin{align}
\hat{H} &= \alpha (\hn_1 + \hn_2) - t \sum_{\sigma} (\had_{1\sigma} \ha_{2\sigma} + \had_{2\sigma} \ha_{1\sigma}) \nonumber \\
&+ U ( \hn_{1\uparrow} \hn_{1 \downarrow} + \hn_{2\downarrow} \hn_{2 \uparrow})  + w \hn_1 \hn_2
\label{eq:Ham}
\end{align}
$\had_{1\sigma}$ ($\ha_{1\sigma}$) are creation (annihilation) operators, $\nh_{i\sigma}= \had_{i\sigma} \ha_{i\sigma}$
and $\nh_i=\nh_{i\uparrow} + \nh_{i\downarrow}$ are the density operators, while the four parameters $\alpha$, $t$, $U$ and $w$ are respectively the on-site and inter-site interactions.
These parameters can be given a dependence on the bond distance $R$ in such a way that the model's bonding curve closely reproduces the real H$_2$ dissociation curve. The proper dependencies were already derived in \cite{Gie-EPJB-18}, using two-electron integrals in a minimal basis of Löwdin orthogonalized atomic Slater functions from the real-space Hamiltonian of the hydrogen molecule; here, we follow the same prescription and our results are presented in Fig.\ref{Hubbard-parameters}, while further details on their derivation can be found in Appendix \ref{Hubbard_parameters}. 
Defining the total number operator as: 
\be
\hN = \hn_1 + \hn_2,
\ee
it can be readily shown that the Hamiltonian in Eq.(\ref{eq:Ham}) can be recast as follows:

\be
\hH = \alpha \hN + \frac{w}{2} ( \hN^2-\hN) + \hH^\prime,
\ee
with:
\be
\hH^\prime = - t \sum_{\sigma} (\had_{1\sigma} \ha_{2\sigma} + \had_{2\sigma} \ha_{1\sigma}) + (U-w) ( \hn_{1\uparrow} \hn_{2 \downarrow} + \hn_{1\downarrow} \hn_{2 \uparrow}) 
\label{Eq:hprime}
\ee
and where $\hH^\prime$ is the Hamiltonian of a standard Hubbard dimer with a renormalized on-site repulsion $U-w$.
Since we will study the 2-particle case, and the Hamiltonian commutes with $\hN$, it follows that the eigenstates of $\hH$ and $\hH'$ are the same,
while the eigenenergies $E_j$ and $E_j^\prime$ are related in the following way:
\be
E_j = 2\alpha + w + E_j^\prime.
\ee
Thus, it suffices to perform all the calculations for $\hH^\prime$ and simply add the quantity $2\alpha+w$ to the final result.
The ground state energy of $\hH^\prime$ can be determined by exact diagonalization and it reads:
\be
E_0^\prime = \frac{1}{2} (U-w) - \sqrt{4t^2 + (\frac{U-w}{2})^2}
\label{eq:E-exact}
\ee
where the energy of the original Hamiltonian (\ref{eq:Ham}) reads:
\be
E_0 = 2 \alpha + w + E_0^\prime.
\ee
The bonding curve for the H molecule at a given bond distance $R$, is then obtained by adding to the ground state energy the term $1/R$, which
accounts for the nuclear-nuclear repulsion. 

\begin{figure}
\includegraphics[width=0.5\textwidth]{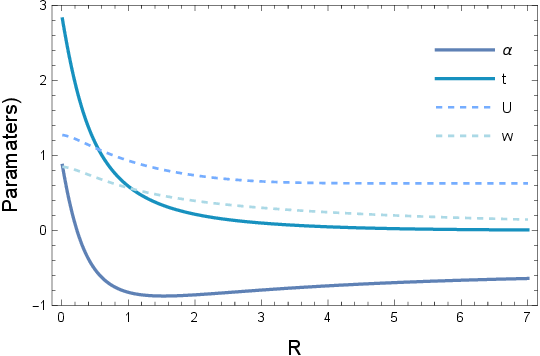} 
\caption{A plot of the Hubbard parameters $\alpha,t,w,U$ (respectively in solid dark and medium blue and in dashed light and very ligth blue) as function of the bond distance $R$.}
\label{Hubbard-parameters}
\end{figure}


\subsubsection{The exact Green's function}
\label{exactG}

The exact Green's function of the Hamiltonian $\hH^\prime$ can be evaluated directly from the Lehmann representation. In the HF basis it has the following expression:
\begin{align}
G_{11} (i \omega) &= \frac{I_+ (\beta)}{i \omega - \Omega_1+ \mu}  + \frac{I_- (\beta)}{i \omega - \bar{\Omega}_2+ \mu}  \\
G_{22} (i \omega) &= \frac{I_- (\beta)}{i \omega - \Omega_2+ \mu}  + \frac{I_+ (\beta)}{i \omega - \bar{\Omega}_1+ \mu}  
\end{align}
where the off-diagonal elements are zero. 
Here $\beta=(U-w)/4t$ and 
\be
I_\pm (\beta) = \frac{1}{2} \frac{\sqrt{1+\beta^2} \pm 1}{\sqrt{1+\beta^2}}
\label{eq:intensity}
\ee
are the pole strengths. The poles are instead given by:
\begin{align}
\Omega_1 &= t + \frac{1}{2} (U-w) - \delta \\
\Omega_2 &= -t + \frac{1}{2} (U-w) - \delta \\
\bar{\Omega}_1 &= -t + \frac{1}{2} (U-w) + \delta \\
\bar{\Omega}_2 &= t + \frac{1}{2} (U-w) + \delta 
\end{align}
where we defined:
\be
\delta = \sqrt{4 t^2 + (\frac{U-w}{2})^2} > 2t 
\ee
The poles $\Omega_1$ and $\Omega_2$ correspond to removal energies while 
$\bar{\Omega}_1$ and $\bar{\Omega}_2$ correspond to addition energies. Setting $\beta=0$ in Eq.(\ref{eq:intensity}) and $\delta=0$ in the equations
for the poles, yields the HF Green's function. 
The above expressions will be used to construct a model input Green's function by setting $\beta=0$ in Eq.(\ref{eq:intensity}), while still keeping $\delta$ in its given form in the equation of the poles. This produces an improved Green's function with an identical structure of the non-interacting one, while at the same time displaying the correct dominant poles.


\subsection{Analytical expressions for the Klein functional}

Evaluating the expression (\ref{eq:energy_Klein}) for the dimer yields a rather compact expression for the Klein functional, namely:
\be
E_\KL [ \tG ] = -2t + \Phi [\tG]
\ee

where for the poles we used that $\epsilon_{1,\HF} = - t + (U-w)/2$ and $\Sigma_{11,\HF} =(U-w)/2$.
More in general, in order to plot the bonding curve $\mathcal{E}(R)$, it is necessary to move from the $\hH^\prime$ of Eq.~\ref{Eq:hprime} back to the original Hamiltonian $\hH$, and this is achieved by evaluating the expression:

\be
\mathcal{E}(R)= 2\alpha  + w + E [\tG] + \frac{1}{R}
\ee
where the term $1/R$ is added to take into account the nuclear-nuclear contribution to the total energy, $E [\tG]$ is either the Klein or the LW functional, and all the terms are functions of $R$ (see the discussion in the section \ref{extended_dimer}).
In this work two different choices are made for $\tG$. 
\begin{enumerate}
\item The first one is $\tG=G_\HF$. In this case we have for the poles that: 
\begin{align}
\epsilon_1 &= \epsilon_{\HF,1} = -t + \frac{U-w}{2}  \\
\epsilon_2 &= \epsilon_{\HF,2} = t + \frac{U-w}{2} 
\end{align}
with the HOMO-LUMO gap $\Delta = 2t$.

\item The second choice amounts to constructing a model Green's function ($G_{\textrm{mod}}$) with poles of unit strength at the position of the dominant poles of the exact Green's function (calculated analytically earlier).
In the language of the poles used in the section \ref{exactG}, 
one takes $\epsilon_1 = \Omega_1$ and $\epsilon_2 = \bar{\Omega}_1$ corresponding to the poles with the dominant strength $I_+ (\beta)$, while omitting all together the secondary shake-up poles of the exact Green's function with weaker intensity $I_- (\beta)$.

This physical idea translates in the following expressions:
\begin{align}
\epsilon_1 &= t + \frac{1}{2}(U-w) - \delta = \epsilon_{\HF,2} - \delta < \epsilon_{\HF,1}  \\
\epsilon_2 &= -t + \frac{1}{2}(U-w) + \delta = \epsilon_{\HF,1} + \delta > \epsilon_{\HF,2} \\
\delta &= \sqrt{4 t^2 + (\frac{U-w}{2})^2} > 2t 
\end{align}

The ordering of the eigenvalues $\epsilon_1 < \epsilon_{\HF,1} <  \epsilon_{\HF,2} < \epsilon_2$ is such that one can find a common chemical potential $\mu$
inside the interval $( \epsilon_{\HF,1},\epsilon_{\HF,2} )$
that gives equal particle number for the HF and model Green's function: this choice is made for all our derivations.
\end{enumerate}

\begin{figure*}[htp]

\begin{subfigure}[b]{0.30\textwidth}
\includegraphics[width=\textwidth]{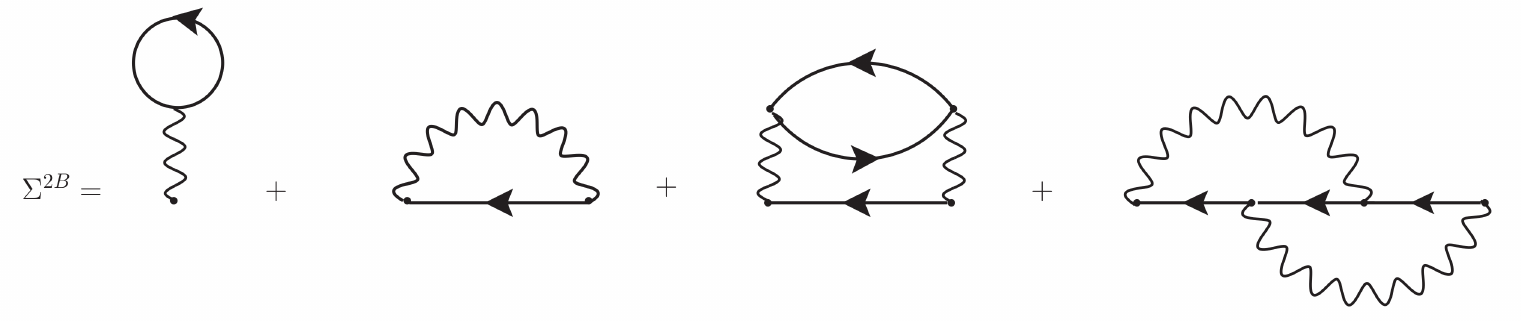}
\caption{2B}
         \label{fig:a}
  \end{subfigure}
  \hfill
  \begin{subfigure}[b]{0.30\textwidth}
\includegraphics[width=\textwidth]{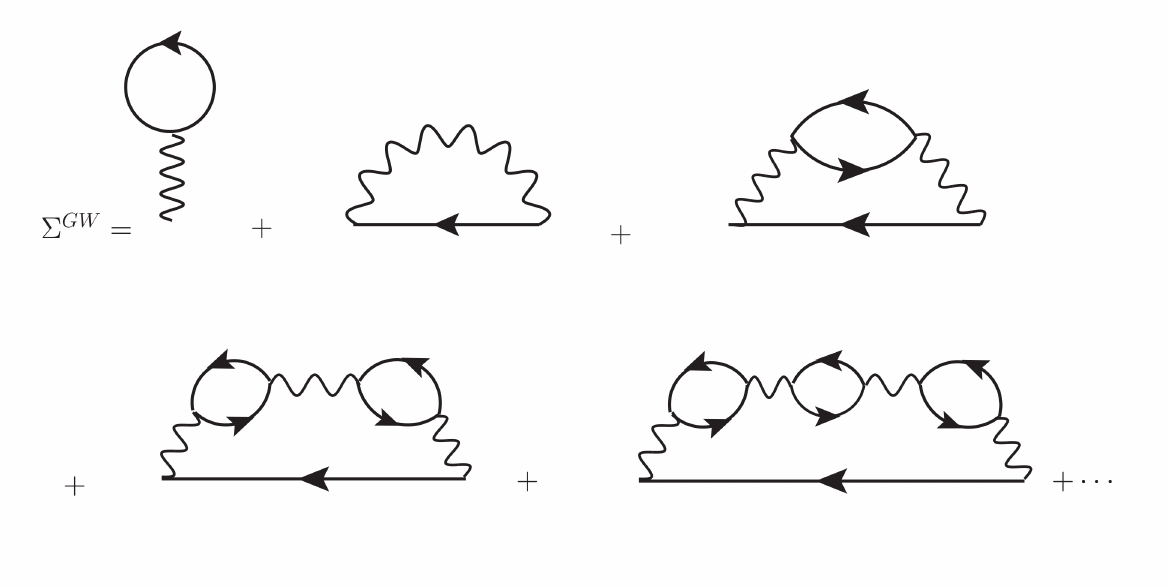}
\caption{GW}
         \label{fig:b}
  \end{subfigure}
\hfill
 \begin{subfigure}[b]{0.30\textwidth}
\includegraphics[width=\textwidth]{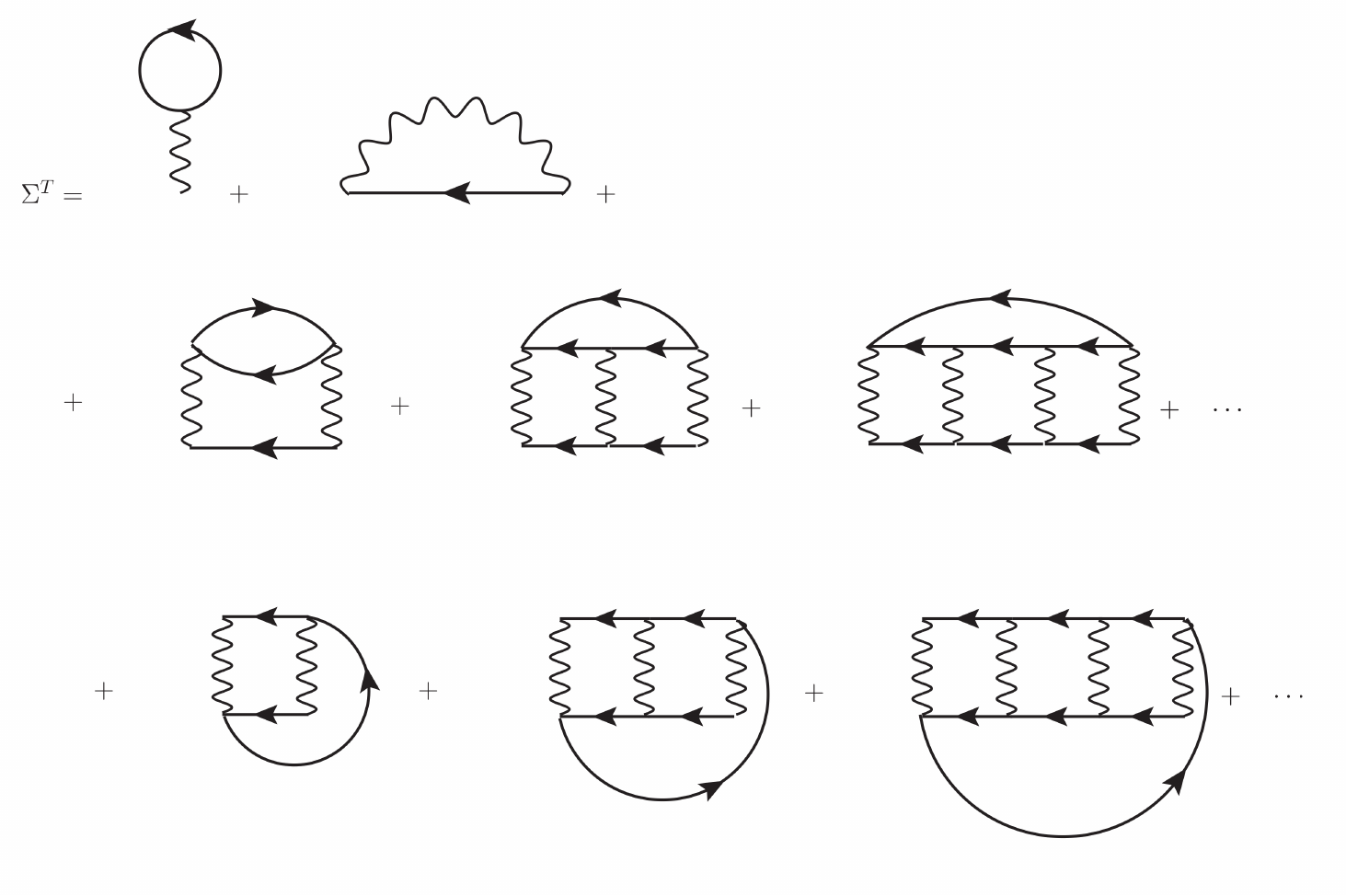}
\caption{T-matrix}
         \label{fig:c}
  \end{subfigure}
\caption{Self-energy diagrams for three selected many-body approximations}
\label{sigma}
\end{figure*}

\begin{figure*}[htp]

\begin{subfigure}[b]{0.30\textwidth}
\includegraphics[width=\textwidth]{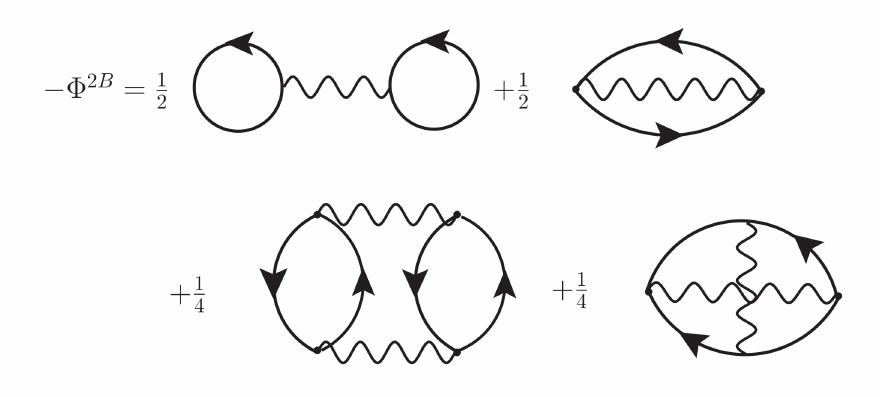}
\caption{2B}
         \label{fig:a}
  \end{subfigure}
  \hfill
  \begin{subfigure}[b]{0.30\textwidth}
\includegraphics[width=\textwidth]{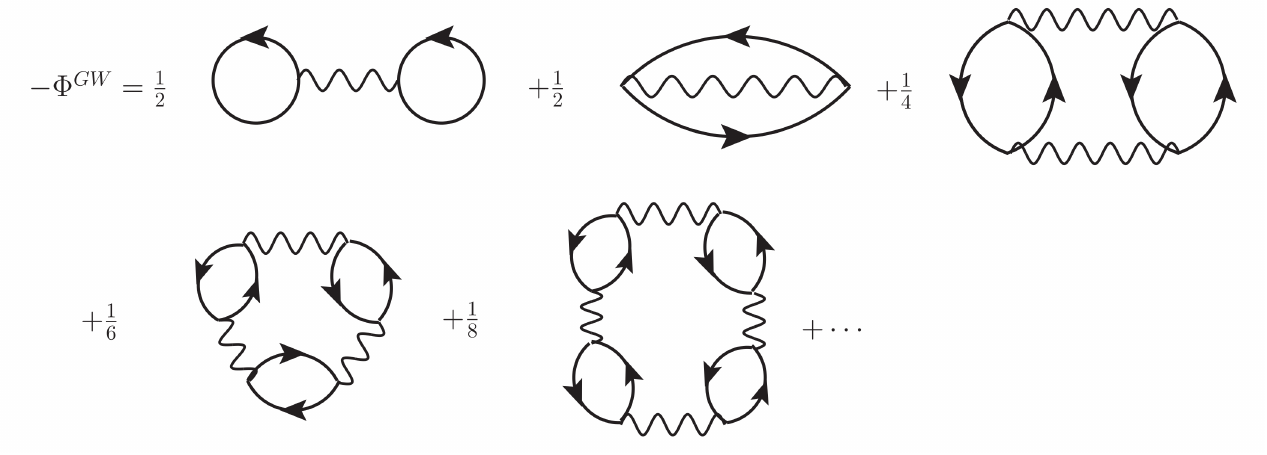}
\caption{GW}
         \label{fig:b}
  \end{subfigure}
\hfill
 \begin{subfigure}[b]{0.30\textwidth}
\includegraphics[width=\textwidth]{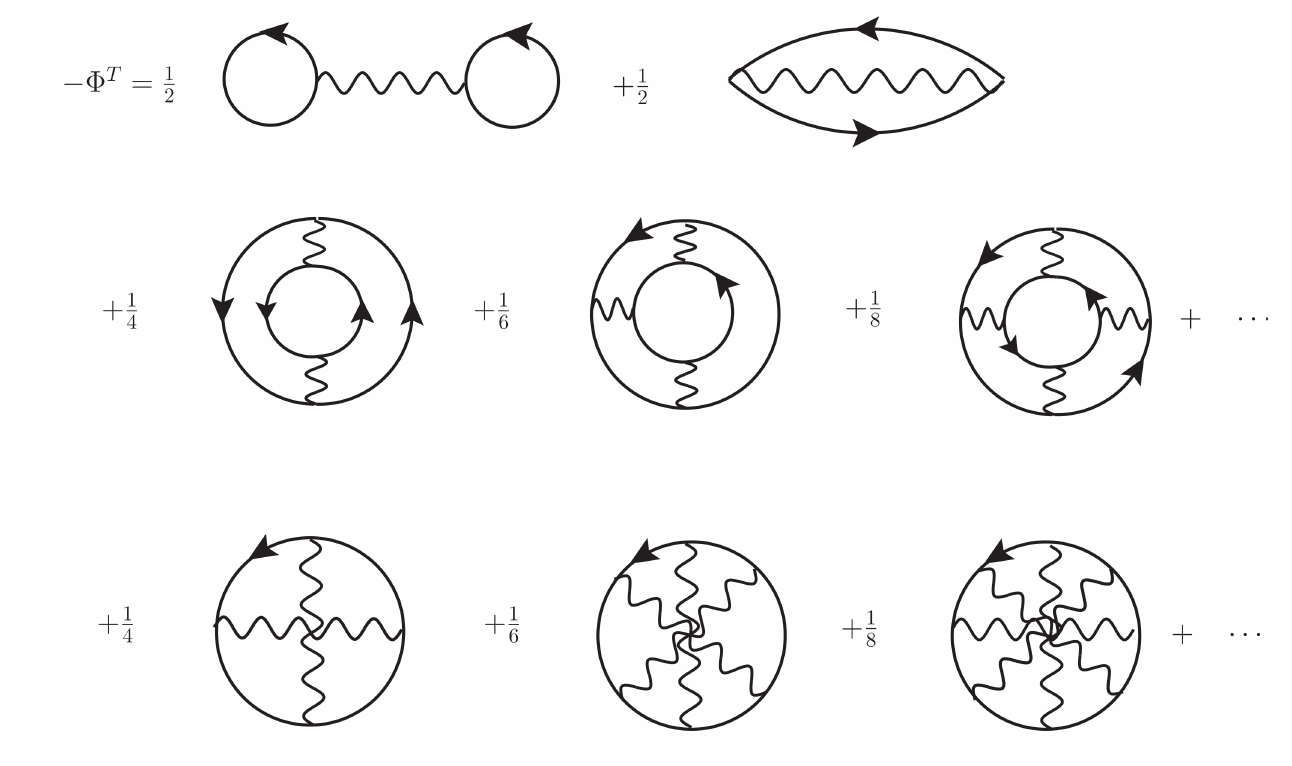}
\caption{T-matrix}
         \label{fig:c}
  \end{subfigure}
\caption{$\Phi$ diagrams for three selected many-body approximations}
\label{phi}
\end{figure*}

We can now begin the study of three selected many-body approximations, that second Born (2B), GW and the so-called T-matrix. The Feynman diagrams for $\Phi$ and $\Sigma$ for the three approximations are shown in Fig.~\ref{sigma} and~\ref{phi}.

The HOMO-LUMO gap reads:
\be
\Delta =\epsilon_2-\epsilon_1
\ee
where for the HF input case $\Delta =2t$ and for the model Green's function $\Delta = 2 \delta - 2t$.
After some algebra, the $\Phi$ functional for the various approximations can be written solely in terms of the gap $\Delta$ and the $U$ and $w$ parameters:

\begin{align}
\Phi_\HF [\tG] &= \frac{1}{2} (U-w) \\
\Phi_\SB [\tG] &= \frac{1}{2} (U-w)  - \frac{1}{8} \frac{(U-w)^2}{\Delta} \\
\Phi_\GW [\tG] &=  \frac{\Delta}{2} ( \sqrt{1 + 2 \frac{(U-w)}{\Delta}} -1) \\
\Phi_\Tm [\tG] &=  \Delta ( \sqrt{1 + \frac{(U-w)}{\Delta}} -1) 
\end{align}

Whenever $U-w < \Delta$, both $\Phi_\GW$ and $\Phi_\Tm$ can be expanded in powers of $(U-w)/\Delta$, yielding:

\begin{align}
\Phi_\GW [\tG] = \frac{1}{2} (U-w)  - \frac{1}{4} \frac{(U-w)^2}{\Delta} + \ldots\\
\Phi_\Tm [\tG] = \frac{1}{2} (U-w)  - \frac{1}{8} \frac{(U-w)^2}{\Delta} +\ldots
\label{Eq:Tmat-correct}
\end{align}

Interestingly, $\Phi_\Tm$ gives the exact expansion up from Eq.(\ref{eq:E-exact}) to second order. This can be understood from the fact that the
diagrams for the T-matrix approximation contain the second Born diagrams (which also happen to provide the expansion of the exact solution up to second order). 
In the GW approximation instead the second order term is way too large (by a factor of two), the reason being that second order GW self-energy contains the direct second order bubble diagram but omits the second order exchange term. 

\begin{figure}
\includegraphics[width=0.5\textwidth]{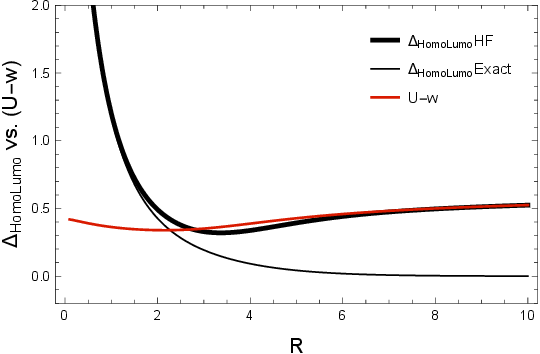} 
\caption{The exact gap $\Delta_\textrm{Exact}$ (thick solid black line) and the HF one (thin solid black line $\Delta_\HF=2t$) used in the model $\tG$ as function of the bond distance $R$ versus
the effective Hubbard interaction $U-w$ (solid red).}
\label{Gap}
\end{figure}

Given the mathematical structure of the above equations, it is instructive to examine how the HF and the exact gap compare to one another and to the quantity $U-w$. This is shown in Fig.\ref{Gap}. 
It can be seen that $U-w$ is smaller than both
the HF or the exact gap up to $R=2$. 
Beyond that range, perturbation theory is formally not applicable. 
However, it can be observed that the exact gap still
remains finite and becomes equal to $U-w$ at large bond distances while the HF gap closes exponentially fast to zero. 
This implies that the HF Green's function becomes a very poor approximation to the true Green's function whenever $R >2$.

\subsection{Analytical expressions for the Luttinger-Ward functional}

 The total energy given by the LW functional can be obtained via the correction given in Eq.(\ref{eq:LW_correction3}), which in turn requires the evaluation of the correlation part of the self-energy $\Sigma_\co [\tG]$. 
 As it turns out, for all the three many-body approximations considered, all the self-energy expressions share a similar structure, which can be summarized as:
 
\begin{enumerate}
\item Second Born (2B):
\begin{align}
 \Sigma_{\co,11}^\SB [\tG] (i \omega) &= \frac{\gamma^\SB}{i\omega -  \epsilon_2 - \Omega^\SB+ \mu}\\
 \Sigma_{\co,22}^\SB [\tG] (i \omega) &= \frac{\gamma^\SB}{i\omega - \epsilon_1 +  \Omega^\SB + \mu}\\
 \gamma^\SB &= \frac{1}{4} (U-w)^2\\
 \Omega^\SB &=\Delta = \epsilon_2-\epsilon_1
\end{align}
\item GW:
\begin{align}
 \Sigma_{\co,11}^\GW [\tG] (i \omega) &= \frac{\gamma^\GW}{i\omega - \epsilon_2 -\Omega^\GW + \mu}\\
 \Sigma_{\co,22}^\GW [\tG] (i \omega) &= \frac{\gamma^\GW}{i\omega - \epsilon_1 +  \Omega^\GW + \mu}\\
 \gamma^\GW &= \frac{\Delta}{2 \Omega^{\GW}} (U-w)^2 \\
 \Omega^\GW &=\sqrt{ \Delta^2+ 2 \Delta (U-w)}
\end{align}
\item T-matrix:
\begin{align}
 \Sigma_{\co,11}^\Tm [\tG] (i \omega) &= \frac{\gamma^\Tm}{i\omega - \epsilon_2 -\Omega^\Tm + \mu}\\
 \Sigma_{\co,22}^\Tm [\tG] (i \omega) &= \frac{\gamma^\Tm}{i\omega - \epsilon_1 + \Omega^\Tm + \mu}\\
 \gamma^\Tm &= \frac{\Delta}{4 \Omega^\Tm} (U-w)^2 \\
 \Omega^\Tm &=\sqrt{ \Delta^2+ \Delta (U-w)}
\end{align}
where the gap $\Delta =\epsilon_2-\epsilon_1$ has been defined as in the preceding section.
\end{enumerate}

In summary, the self-energy for the three approximations can be cast in the following general form:
\begin{align}
 \Sigma_{\co,11}^\mx [\tG] (i \omega) &= \frac{\gamma^\mx}{i\omega -  \epsilon_2 - \Omega^\mx+ \mu}\\
 \Sigma_{\co,22}^\mx [\tG] (i \omega) &= \frac{\gamma^\mx}{i\omega - \epsilon_1 +  \Omega^\mx + \mu}
 \end{align}
where $\mx=\SB,\GW,$ and $T$, and the parameters $\Omega^\mx$ and $\gamma^\mx$ are chosen accordingly to the equations specified above. 
This observation and choice of notation greatly simplifies the evaluation of the LW correction, since it acquires the following form (irrespective of the many-body approximation considered):

\be
C_\LW^\mx [\tG] =   - \Tr (\tG \Sigma_\co^\mx [\tG])  - \Tr \ln (1-G_\HF \Sigma_\co^\mx [\tG]) .
\label{Eq:lw-corr-m}
\ee

For the first term of Eq.\ref{Eq:lw-corr-m}, we find by direct evaluation of residue integrals that:

\be
- \Tr (\tG \Sigma_\co^\mx [\tG]) = \frac{4 \gamma^\mx}{\Delta + \Omega^\mx},
\ee
while for the second term we can employ the result (\ref{eq:Lambda_int2}) derived in the Appendix \ref{app:logint} to find:

\begin{align}
- \Tr &\ln (1-G_\HF \Sigma_\co^\mx [\tG]) = \nonumber \\
&- \frac{4 \gamma^\mx}{| \epsilon_{\HF,1}-\epsilon_2 - \Omega^\mx| + 
\sqrt{ (\epsilon_{\HF,1}-\epsilon_2 - \Omega^\mx)^2 + 4\gamma^\mx}}  \nonumber\\
&\quad - \frac{4 \gamma^\mx}{| \epsilon_{\HF,2}-\epsilon_1 + \Omega^\mx| + 
\sqrt{ (\epsilon_{\HF,2}-\epsilon_1 + \Omega^\mx)^2 + 4\gamma^\mx}} 
\end{align}

The poles $\epsilon_1$ and $\epsilon_2$ are chosen depending on which input Green's function (HF or model one) is used.
For the HF Green's function one has:

\be
C_\LW^\mx [G_\HF] = \frac{4 \gamma^\mx}{\Delta + \Omega^\mx} - \frac{8 \gamma^\mx}{\Delta + \Omega^\mx + \sqrt{(\Delta + \Omega^\mx)^2 + 4 \gamma^\mx}} 
\label{Eq:clwghf}
\ee

and where the quantities $\Omega^\mx$ and $\gamma^\mx$ should be evaluated using the specified gap $\Delta=2t$. 
It is readily seen from this expression that $C_\LW^\mx [G_\HF] >0$ and therefore the LW correction gives a positive shift to the energy.\\

In the case of the model Green's function $G_{\textrm{mod}}$ one has:
\begin{align}
C_\LW^\mx [G_{\textrm{mod}}] &= \frac{4 \gamma^\mx}{\Delta + \Omega^\mx} - \frac{8 \gamma^\mx}{ \delta + \Omega^\mx + \sqrt{(\delta+ \Omega^\mx)^2 + 4 \gamma^\mx}} 
\label{eq:tG_correction}
\end{align}
where the quantities $\Omega^\mx$ and $\gamma^\mx$ should be evaluated using a different gap, that is $\Delta=2 \delta-2t$, where $\delta = \sqrt{4 t^2 + (\frac{U-w}{2})^2} \,> \,2t$.

In this latter case, the LW correction is negative, $C_\LW^\mx [G_{\textrm{mod}}] <0$, for both the 2B and T-matrix approximations, as it is shown in Appendix \ref{LW_model}.  
For the GW approximation instead, a positive shift is theoretically possible, although for the H$_2$ model its value would be very small and only for short bond distances; therefore, also for this case the negative corrections have a much larger contribution in most of the bonding range.


\section{Results for the molecular model}

In this section we present our results for two different Green's function inputs, namely the HF and the model one, combined with three many-body approximations, specifically 2B, GW and T-matrix.\\
We shall begin with some general considerations and then dive into the specificity of the two functional with various combinations of the main ingredients.

\begin{figure}[ht]
  \centering
\subfloat[2B approximation with a Hartree-Fock Green's function as input]{
   \includegraphics[width=0.45\textwidth]{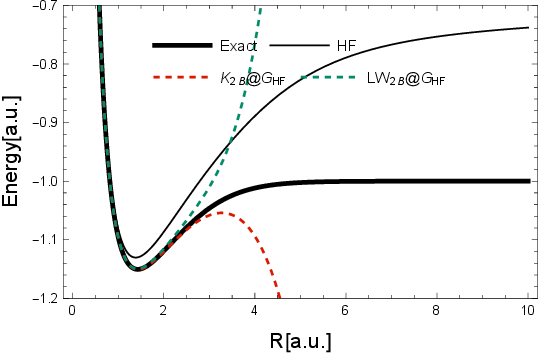}
   \label{fig:f1}
  }
  \hfill
  \subfloat[2B approximation with the model Green's function as input]{
    \includegraphics[width=0.45\textwidth]{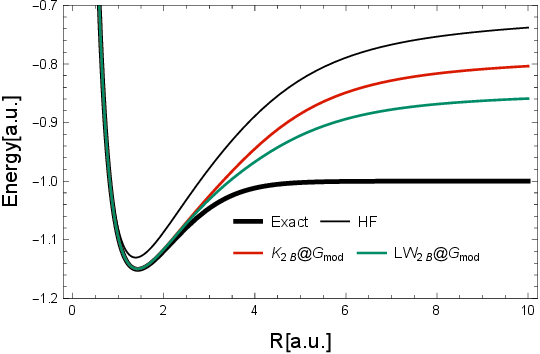}
    \label{fig:f2}
  }
\caption{Potential energy surfaces for both the Klein and the LW functionals, at the 2B level, using $G_{\HF}$ (left panel) and the model $G_{\textrm{mod}}$ (right panel). When using $G_{\textrm{mod}}$, both the Klein functional curve (solid red) and the LW one (solid dark green) lie between the exact curve (solid thick black) and the HF one (solid thin black). When using HF input, both the Klein (dashed red) and the LW curves (dashed dark green) exhibit a divergent behaviour for $R > 3$. 
}
  \label{Fig_2B}
\end{figure}

\begin{figure}[h]
  \centering
  \subfloat[GW approximation with a Hartree-Fock Green's function as input ]{\includegraphics[width=0.45\textwidth]{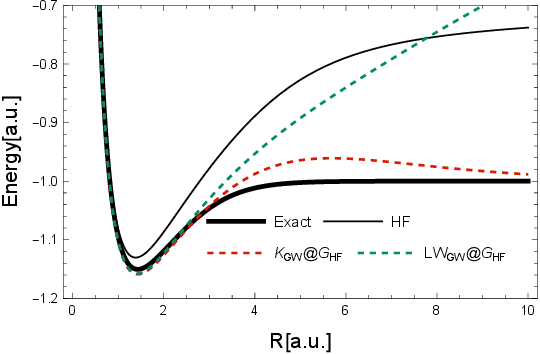}\label{fig:f3}}
  \hfill
  \subfloat[GW approximation with the model Green's function as input]{\includegraphics[width=0.45\textwidth]{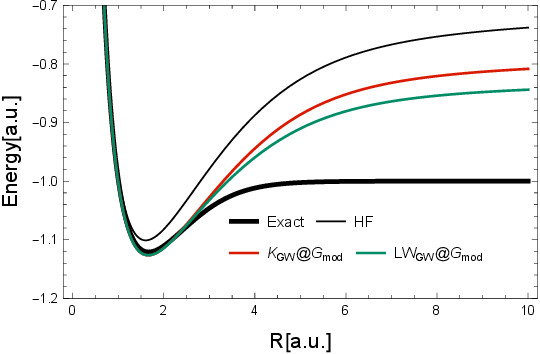}
  \label{fig:f4}}
  \caption{Potential energy surfaces at the GW level for the HF input (left panel) and the model $G_{\textrm{mod}}$ input (right panel). The HF (thin solid black line) and the exact curve (thick solid black line) are also provided for comparison. When using $G_{\textrm{mod}}$, both the Klein functional curve (solid red) and the LW one (solid dark green) lie between the reference curves, just like for the 2B approximation.
  Conversely, in the case of the $G_{\HF}$ input, the LW curve (dashed dark green) crosses the HF one for larger R values, while the Klein curve (dashed red) approaches the correct dissociation limit.
  }
  \label{Fig_GW}
\end{figure}

\begin{figure}[hb]
  \centering
  \subfloat[T-matrix approximation with a Hartree-Fock Green's function as input]{\includegraphics[width=0.45\textwidth]{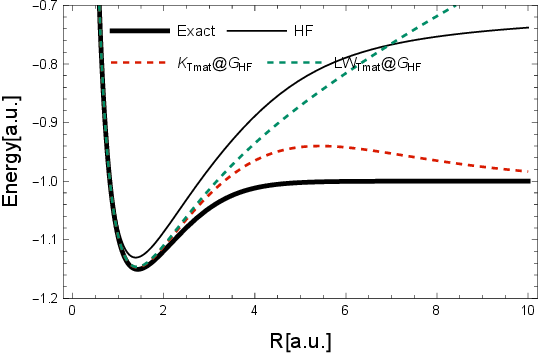}\label{fig:f5}}
  \hfill
  \subfloat[T-matrix approximation with the model Green's function as input]{\includegraphics[width=0.45\textwidth]{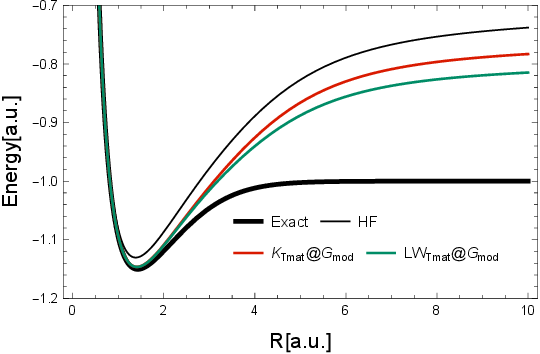}\label{fig:f6}}
  \caption{Potential energy surfaces at the T-matrix level for the HF input (left panel) and the model $G_{\textrm{mod}}$ input (right panel). The HF (thin solid black line) and the exact bonding curve (thick solid black line) are also provided for comparison.
  Once again, when using $G_{\textrm{mod}}$, both curves (Klein functional in solid red and LW solid dark green) are sitting between the reference ones. Using the HF Green's function, drastically worsens the LW results (dashed dark green) while again the Klein curve (dashed red) reaches the dissociation limit accurately 
  }
    \label{Fig_Tmat}
\end{figure}

The general trend is the same for all many-body approximations; however, some important differences, which we will thoroughly address, appear.
Up until a bond distance of $R \approx 3$ (which includes the equilibrium value), employing both the HF and model Green's function in conjunction with any of the three many-body approximations leads to results that are all fairly close to each other and to the exact bonding curve. This success can be attributed to the fact that in this particular regime the exact and the HF gap are still comparable (see once again Fig.~\ref{Gap}). 
This is shown in both panels of Fig.~\ref{Fig_2B} for the 2B approximation, as well as in Figs.~\ref{Fig_GW} and ~\ref{Fig_Tmat} for the GW and the T-matrix.
For $R > 3$, employing the HF Green's function, the scenario is remarkably different, can be written solely be attributed to the poor representation of the HOMO-LUMO gap by the HF Green's function (see \ref{Gap}).
The energy curves calculated with the Klein and the LW functional deviate considerably from one another, see in particular Figs.\ref{fig:f3} and \ref{fig:f5}: in both cases the LW functional curve grows monotonically and even crosses the HF result, reaching though a finite value in the dissociation limit (this was also verified analytically).
The large discrepancy between the Klein and the LW results is a clear indication that the input Green's function is rather removed from that of a self-consistent solution of the Dyson equation. This implies that the second-order correction in $\Delta G$ in  Eq.(\ref{DeltaG_exp}) is not small, and thus we are evaluating the functionals far from their stationary point. \\

\begin{figure}[h]
\includegraphics[width=0.45\textwidth]{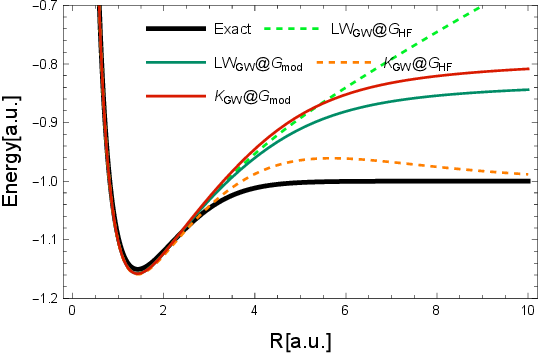} 
\caption{Potential energy surfaces computed with the LW and the Klein functional at the GW level with $G_{\textrm{mod}}$ and $G_{\HF}$ (in solid and dashed dark green respectively) versus the energies obtained with the Klein one, in solid red and orange dashed (respectively for $G_{\textrm{mod}}$ and $G_{\HF}$ input). The exact energy (thick solid black) is also reported for comparison.}
\label{LW_stability}
\end{figure}

Conversely, when using the model Green's function, whose distinctive feature is having a HOMO-LUMO gap resembling that of the exact Green's function, it can be observed in Figs.\ref{fig:f2}, \ref{fig:f4} and \ref{fig:f6} that $E_\KL [G_{\textrm{mod}}] \approx E_\LW [G_{\textrm{mod}}]$ for all the three many-body approximations. In all cases, the LW functional improves the results obtained with the Klein one (as expected) and even cures the divergence observed in Fig.~\ref{fig:f2} for the case of the 2B approximation.
The fact that $E_\KL [G_{\textrm{mod}}] \approx E_\LW [G_{\textrm{mod}}]$ also implies that the LW bonding curves for the model input are expected to be rather close (at least from the theoretical standpoint) to those that one would obtain from a self-consistent calculation. 
This is indeed confirmed by a fully self-consistent numerical solution of the Dyson equation, as will be discussed in much more detail below. So also from the numerical standpoint we conclude that $E_\KL [G_{\textrm{mod}}] \approx E_\LW [G_{\textrm{mod}}] \approx E_{sc}[G]$.\\
Our numerical results for the case of 2B are further supported by Ref.~\cite{Dah-PRA-06}; in Fig. 4 of the work, the 2B curve calculated fully self-consistently is seen to be very close to the one obtained in our work. It sits below the HF one and reaches a finite value at large R, albeit higher than the exact limit. It is also worth noting that for the 2B approximation the self-consistency completely removes the divergence, something which was also pointed out in \cite{RusZgi-JCP-16}.
This evidence makes the 2B approximation rather appealing for quantum chemistry calculations, as it is computationally inexpensive, yet accurate at various correlation regimes (barring the dissociation one, where in any case also more sophisticated many-body approximations exhibit shortcomings). \\
Another important general trend observed in the calculations is the overall better stability of the LW functional compared to the Klein one: in other words, the results obtained using the LW functional are less sensitive to the quality of the input Green's function. For example, in Fig.\ref{LW_stability}, for the GW approximation, it can be seen that $E_\LW [G_{\textrm{mod}}]$ and $E_\LW [G_\HF]$ are very close up to $R \approx 5$, while for the Klein functional the energy curves are in agreement only up until $R\approx 2$ and then start to move apart. In \cite{Sta-JCP-09}, the self-consistent $GW$ result and the $E_\LW[G_\HF]$ for the H$_2$ molecule were found to be very close even up to $R=7$. Those calculations were performed with a much larger basis than the model system; nonetheless, the conclusions that can be drawn are just the same.
The remarkable stability of the LW functional was also pointed out in another series of earlier works: in \cite{Dah-PRA-06} and \cite{Dah-JPC-05}, respectively in Fig. 4 and Fig. 3, it was noted that the energy calculated from the Galitskii-Migdal formula (in conjunction with a self-consistent calculation) and that from the LW functional (with a HF input) are very close up to $R=7$ for the 2B approximation. Other self-consistent results for 2B can be found in \cite{RusZgi-JCP-16} for an infinite chain of hydrogen atoms, and in \cite{PhiZgi-JCP-14} for various H$_n$ molecules, and they all exhibit the trend highlighted above. \\

\begin{figure}
\includegraphics[width=0.45\textwidth]{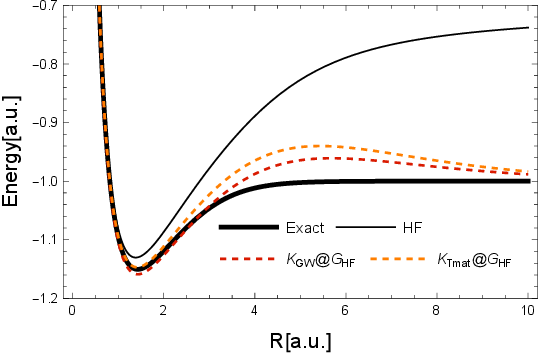} 
\caption{Potential energy surfaces computed with the Klein functional with
$G_\HF$ as an input at the GW (red dashed) - equivalent to the RPA - and at the T-matrix level (orange dashed). 
The exact curve (thick solid black) and Hartree-Fock one (thin solid black) are also shown for comparison. }
\label{Fig:rpa-comparison}
\end{figure}

For the Klein functional with $G_{\HF}$ as an input, the bonding curves for both GW and T-matrix happen to be close to the exact results apart from a "bump" \cite{Fuc-PRB-02} at around $R=5$ (see Fig.\ref{Fig:rpa-comparison}), which is a very well known behavior of the random phase approximation when used within the adiabatic-connection fluctuation-dissipation (ACFD) framework. These results deserve a separate discussion.
The ACFD framework within DFT has been considered for quite some time a promising tool for inexpensive albeit rather accurate calculations of correlation energies \cite{Fuc-PRB-02, HelBar-PRB-07, HelBar-PRB-08, Ngu-PRB-09, TouGerJan-PRL-09, Har-PRL-09, Hes-PRL-11, Ren-PRL-11, OlsThy-JCP-14, Hel-JCP-23, Agg-JCP-14}. 
In a number of different works~\cite{Hell-PRB-15,Dah-PRA-06} it has been proven that there is a strict equivalence between the energy calculated with the RPA approximation in the ACFD scheme and that from a Klein functional in conjunction with the GW approximation. 
A notable feature of the energy calculated with the above methodology, or its slightly improved versions \cite{Fuc-JCP-05,Col-PRB-16}, is the accuracy around the dissociation limit for the bonding curve of several atoms and small molecules, including the challenging case of H$_2$.
There are different ways to improve on the ACFD-RPA approach \cite{Gor-PRB-19, Gou-JCP-12}, arguably one of the most systematic ones is the construction of exchange-correlation kernels for time-dependent DFT beyond the exact exchange one \cite{HelBar-PRB-07,HelBar-PRB-08,Hel-PRB-18}. In this latter work, employing a vertex correction based on the exact-exchange potential of TDDFT, the authors obtained some improvements for the H$_2$ dissociation curve over the whole range of bond distances. 
Interestingly, the improvements that we obtained using the Klein functional with the T-matrix approximation and $G_{\HF}$ at the minimum, due to the exact first terms in the expansion of Eq.~\ref{Eq:Tmat-correct},  may have a similar physical justification to those that were obtained in \cite{Hel-PRB-18} following a different route. 

To get a more quantitative perspective, we also compiled a table (see Tab.\ref{tab:table} in App.\ref{app:coeff}) containing the values of the total energy $E_0=E(R_0)$ at the optimised bond distance $R_0$, for both energy functionals with all the approximations considered so far. Additionally, we also reported the harmonic and anharmonic coefficients and their deviation from the exact values (in parenthesis).
It is not straightforward to extract an overall trend. However, at the minimum, the approximation that seems to perform best is the 2B one, no matter which functional or input Green's function is used. This is no longer true for stronger electronic correlation, but it is certainly worth to keep this fact in mind if one is focused on properties related to the energy minimum, such as forces or phonons for which recently an exact $\Phi$-derivable theory was developed \cite{Stefanucci2023}. 

\begin{figure}
\includegraphics[width=0.45\textwidth]{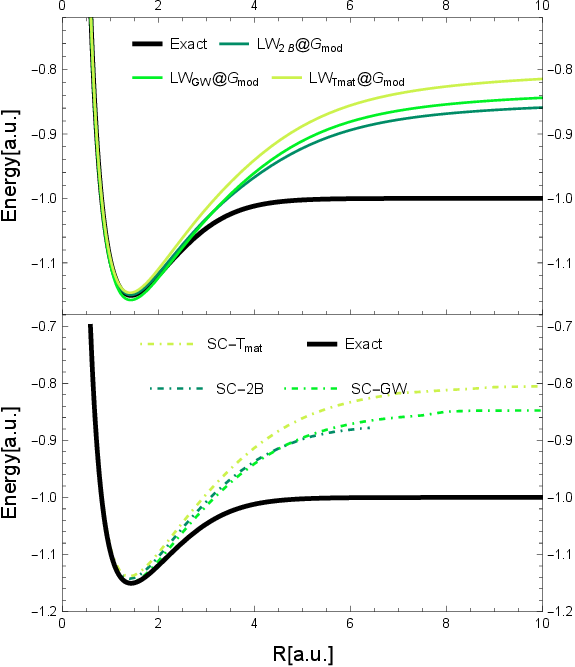} 
\caption{In the top panel, the potential energy surface for the LW functional calculated with $G_{\textrm{mod}}$ and three different many-body approximations, namely 2B, GW and T-matrix (in solid dark, medium and light green respectively) is shown together with the reference exact curve (solid thick black).
In the bottom panel the potential energy surface has been obtained via fully self-consistent (numerical) calculations with the 2B, GW and T-matrix approximations (in dashed-dotted dark, medium and light green). }
\label{LW_comparison}
\end{figure}

\begin{figure}[tb]
\includegraphics[width=0.45\textwidth]{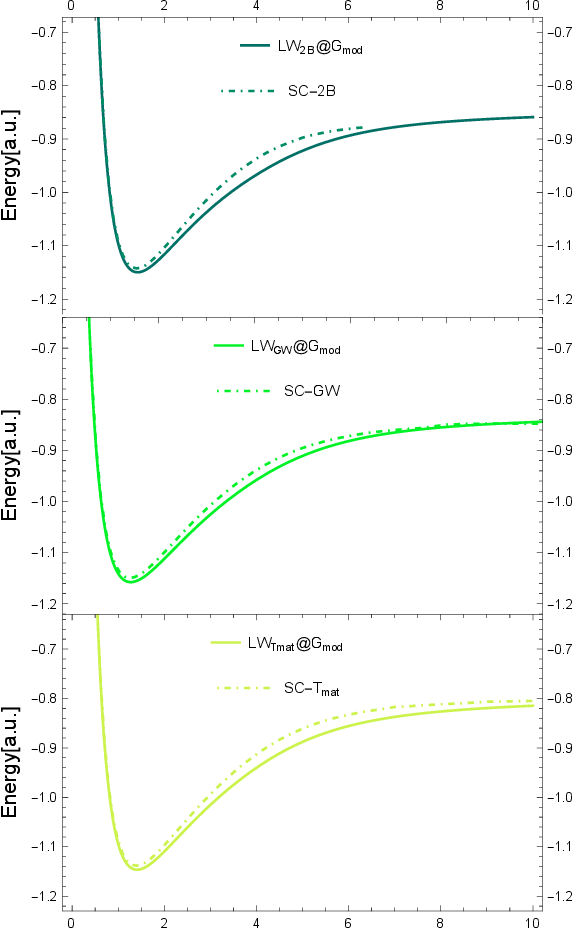} 
\caption{In each panel it is shown the comparison between the potential energy surface obtained through a numerical fully self-consistent calculation, versus that calculated with the LW functional using the model Green's function. The self-consistent results are always depicted with a dash-dotted line, while the LW results always with a solid line. From the top one has the 2B, the GW and the T-matrix approximations. In all cases there is good agreement between the different curves.}
\label{SC-comparison}
\end{figure}

From all of our calculations, it can be noted that the correction term $C^\mx [G_\HF]>0$ is always positive (see Eq.\ref{Eq:clwghf}), making the LW curves lies above the Klein ones for a $G_\HF$ input. On the other hand, $C^\mx [G_{\textrm{mod}}]$ is generally negative, which is strictly proven for 2B and the T-matrix in Appendix \ref{LW_model} and nearly everywhere for the GW case; this makes the LW curves lie below the Klein ones with $G_{\textrm{mod}}$ as input. \\
Finally we compare, in Fig.\ref{LW_comparison} and Fig. \ref{SC-comparison}, the variational functionals to results obtained from a fully self-consistent solution of the Dyson equation, which is a veritable test to determine to which degree the approximate inputs resemble the fully self-consistent ones. The fully self-consistent data were obtained 
from the authors of \cite{Gie-EPJB-18}, which apart from the GW data published in the cited reference also consists of previously unpublished 2B and T-matrix results.
When zooming into the model curves, for $R<1.8$, a very small positive correction $C^\GW [G_{\textrm{mod}}]$ can be observed, however it is negligible compared to the negative corrections when $R > 2$, where it can be clearly seen that the LW curves always lie below the Klein ones. \\
In Fig.\ref{SC-comparison} it can be observed that the LW energy $E_\LW^\mx [G_{\textrm{mod}}]$, calculated with the model G (which gives our "best result)" is essentially on top of the corresponding many-body self-consistent calculation at all three diagrammatic levels, this being particularly true around equilibrium and for larger R values.
There seems to be a slightly better overall agreement for the GW approximation (in the middle panel of the figure), but it's not significant enough to be discussed. The 2B self-consistent curve was obtained only up until $R\approx 6$, due to numerical difficulties in converging the calculation for larger distances, but this does not hinder our conclusions.
Along the same lines, in the top panel of Fig.~\ref{LW_comparison}, we compare the $E_\LW^\mx [G_{\textrm{mod}}]$ curves for the three many-body approximations. Interestingly, the ordering of the curves matches that of the corresponding self-consistent calculations as it can be seen from the bottom panel of the same figure.
This is one of the key results of this work, since it clearly demonstrates how the LW functional, used together with a rather accurate input, is able \textit{to predict} the outcome of fully self-consistent calculations: something of the utmost practical usefulness. It is well known that one of the caveats of many-body perturbation theory is the relatively high computational cost, particularly whenever self-consistent calculations are required. Furthermore, it's not always guaranteed that simply adding more diagrams to an initial set will yield improved results: there are many subtleties that needs to be taken into account~\cite{SteLeeu}. Thus a very effective option to speed up the process would be to employ the LW functional as a tool to preemptively \textit{screen} a set of many-body approximations in order to pick the most accurate one, which can later be used in a full fledged sc-calculation. 
In a recent study~\cite{Wen-JCTC-24} it has been shown that sc-calculations can often be more accurate than those including some sort of vertex correction, when computing molecular ionization potentials. This development further supports the utility of the aforementioned screening procedure. 

\section{Conclusions}

In this work we investigate the performance of two many-body functionals --- Luttinger-Ward and Klein --- in predicting total energies when used in combination with various classes of $\Phi$-derivable, and thus conserving,
many-body approximations.
The choice of performing the study on the challenging case of the extended Hubbard dimer, a paradigmatic and solvable case of molecular dissociation, confers it a fully analytical character, providing a detailed and unprecedented knowledge of the dependence of the functionals on the model parameters, which enables the calculation of special limits and the benchmark of numerical results. Many of our findings bear a conceptual character and are thus general, without being restricted to the model system chosen. \\
A first important result is the confirmation of the improved variational properties of the LW functional over the Klein one, which was already noted in a few earlier numerical works. Furthermore, we show how it is possible to obtain the LW functional from the Klein one by the addition of a compact, elegant and simple to evaluate correction.
The study of the potential energy surface of the extended Hubbard dimer leads to a number of key observations: 2B, GW and the T-matrix are all very accurate at equilibrium and in the presence of a weak correlation regime (corresponding to an inter-site distance of $R=3$) irrespective of the type of input Green's function employed in the calculations. However, at larger distances and at dissociation, this is no longer the case. 
Among the many-body approximation tested, we find that 2B is the most accurate of all over the whole range of inter-site distances. 
This may seem surprising in the first instance, given that the approximation contains a small and finite set of Feynman diagrams, however it includes the exchange ones, which have been shown to be crucial to obtain an accurate description of the properties of finite systems \cite{PhiZgi-JCP-14}. The fact that 2B outperforms the more sophisticated T-matrix, which also includes the exchange diagrams, can be attributed to the fact that in the dimer only a limited number of excitations can occur, and while 2B capture them all, the extra diagrams contained in the T-matrix worsen an already good result.   
Apart from the innate shortcomings of the many-body approximations, the role of the input Green's function becomes crucial: a model G that is constructed to reproduce the HOMO-LUMO gap of the exact one, allowed us to obtain reasonable results for all the three many-body approximations. While the potential energy surfaces remain far from the exact one in the dissociation regime, they are on top of previous numerical self-consistent calculations. This clearly shows how the functionals, and particularly the LW one, used together with a cleverly designed input GF, can yield results as accurate as those of a fully self-consistent calculation, but at a much cheaper cost.
As a corollary, the ordering of the dissociation curves in the one-shot calculation is the same as for the self-consistent ones. This suggests that one could use the LW functional to screen many-body approximations, prior to performing more expensive and accurate fully self-consistent ones. Last but not least, many-body functionals can be systematically improved by adding further diagrams, or can tackle the problem of more general Hamiltonians, as in the case of electron-phonon couplings.
In the light of the present results, we believe that this work would pave the way to explore and adopt more systematically many-body functionals for accurate and inexpensive total energy calculations, and for calculations of other properties of interest in real materials.

\section{Acknowledgements}
The authors gratefully acknowledge Robert van Leeuwen and Klaas Giesbertz for sharing the results of their many-body self-consistent calculations.\\
This research was supported by the NCCR MARVEL, a National Centre of Competence in Research, funded by the Swiss National Science Foundation (grant number 205602).

\appendix

\section{Distance dependence of the Hubbard parameters for the H$_2$-model}
\label{Hubbard_parameters}

\begin{figure}[h]
\includegraphics[width=0.45\textwidth]{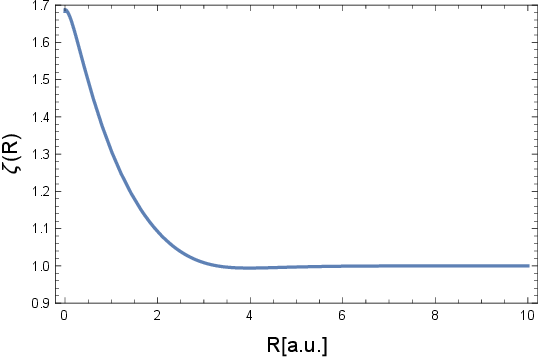}
\caption{Optimized Slater exponent $\zeta(R)$ as a function of the site distance $R$}
\label{zeta}
\end{figure}

We give here the explicit formulas for the $R$-dependence of the parameters $\alpha,t,U,w$ used in the H$_2$ model.
They are derived in \cite{Gie-EPJB-18} from two-electron integrals in a minimal basis of Löwdin orthogonalized atomic Slater functions.
Since we make slightly different choices from the original paper, we give the explicit formulas needed to reproduce the results of this work.
The parameters are given by
\begin{align}
\alpha &= \frac{h_{11} - s \, h_{12}}{1-s^2} \\
t &= -\frac{h_{12} - s \, h_{11}}{1-s^2} \\
U &= I_{1111} + \frac{s^2}{2(1-s^2)} (I_{1111}-I_{1122}) \\
w &=I_{1122} + \frac{s^2}{2(1-s^2)} (I_{1122}-I_{1111}) 
\end{align}
where for $t$ we use the opposite sign of the aforementioned Ref.\cite{Gie-EPJB-18} to be in accordance with the most widespread definition of the Hubbard hopping.
The functions on the right hand side of these equations have the explicit form:
\begin{align}
s &= (1 + \rho + \frac{\rho^2}{3} ) e^{-\rho} \\
h_{11} &= \frac{\zeta^2}{2} -\zeta + \frac{\zeta}{\rho} ( e^{-2\rho} (1+ \rho )-1 )\\
h_{12} &=e^{-\rho} ( \frac{\zeta^2}{2} ( 1 + \rho - \frac{\rho^2}{3}) - 2 \zeta (1+\rho ))\\
I_{1111} &=\frac{5 \zeta}{8} \\
I_{1122} &= \zeta ( \frac{1}{\rho} - e^{-2\rho} (\frac{1}{\rho} + \frac{11}{8} + \frac{3 \rho}{4} +  \frac{\rho^2}{6})
\end{align}
where $\rho =\zeta R$ with $R$ being the bond distance, and $\zeta$ the Slater parameter appearing
in atomic orbitals of the form $\chi(r) =(\zeta^3/\pi)^{1/2} e^{- \zeta r}$. 
In \cite{Gie-EPJB-18}, $\zeta$ is optimized for each bond distance $R$ such that the exact total energy of
the Hubbard system is
minimized, and we repeat the same procedure; the result is shown in Fig.\ref{zeta}.

For $R \rightarrow \infty$, we see that $\zeta (R) \rightarrow 1$, which is to expected as in this limit one obtains two separate hydrogen atoms, which are exactly described
by a Slater function of exponent one. For $R \rightarrow 0$, a helium atom of nuclear charge two is formed, however the electron repulsion reduces this to an effective charge of about $\zeta(0)=1.7$. 

\section{Logarithmic integral}
\label{app:logint}

Most integrals in the main text can be calculated using the Cauchy residue theorem. However, in order to calculate the trace of the logarithmic term for the LW correction in Eq.\ref{Eq:lw-corr-m}, one should be able to perform the following integral:
\be
\Lambda (a,b,c) = \int_{-\infty}^\infty \frac{d\omega}{2 \pi} \ln \left[ 1 - \frac{a}{(i \omega-b)(i \omega -c)}\right],
\ee
where $a,b,c$ are real and $a >0$. The calculation is carried out by first performing a partial integration, followed by successive applications of the Cauchy theorem, together with a careful consideration of the position of the logarithmic branch cut. 
The result of the integration reads:
\begin{align}
\Lambda (a,b,c) = \frac{a}{2 Q} \Big{[} 2 &+ \frac{a-2bc -2 b^2}{|bc-a| + b^2 + |b| Q} \nonumber \\
&+ \frac{a-2bc -2 c^2}{|bc-a| + c^2 + |c| Q} \Big{]},
\label{eq:Lambda_int}
\end{align}
where 
\be
Q(a,b,c) = \sqrt{b^2 +c^2 + 2a + 2 |bc-a|}.
\ee
If the parameters $b$ and $c$ have opposite sign, i.e. one of the values is positive and the other one is negative, then the expression (\ref{eq:Lambda_int}) can be simplified into the more compact form:
\be
\Lambda (a,b,c) = \frac{2a}{|b-c| + \sqrt{(b-c)^2 + 4a}},
\label{eq:Lambda_int2}
\ee 
which is the expression used in the main section of the paper.


\section{Summary of total energy results}
\label{app:coeff}

In this section we present a table with the values of the total energy $E_0=E(R_0)$ at the optimised bond distance $R_0$, i.e. the minimum of the curve, extracted from the bonding curve $E(R)$ for all our approximations. We further define the coefficients
\be
c_n = \frac{d^n E (R)}{d R^n} |_{R_0}
\ee  
where $c_2$ are the harmonic coefficients and $c_n$ for $n \geq 3$ the anharmonic ones.
Since $R_0$ is at the minimum $c_1=0$, we have the Taylor expansion:
\be
E(R) = E_0 + \frac{1}{2}  c_2 (R-R_0)^2 + \frac{1}{6}  c_3 (R-R_0)^3 + \frac{1}{24} c_4 (R-R_0)^4 +  \ldots
\ee
around $R=R_0$. 
Additionally the table lists the values of the harmonic coefficient $c_2$, and the anharmonic coefficients $c_3$ and $c_4$.

The deviation $\Delta x = x_{\textrm{approximation}}- x_\textrm{exact}$ for all of the calculated quantities $x=(E_0,R_0,c_2,c_3, c_4)$ is given in brackets.
The self-consistent results are calculated at too few points to make a reliable estimate of of $c_3$ and $c_4$ possible, thus these coefficients have been omitted from the table.

\begin{table}
\footnotesize
\caption{\label{tab:table}Minimum of the total energy, harmonic and anharmonic coefficients for all the approximations considered in this work}
\begin{ruledtabular}
\begin{tabular}{|c|c|c|c|c|c|}
& $E_0$ & $R_0$ & $c_2$ & $c_3$ & $c_4$ \\
\hline
Exact &-1.15041 & 1.42804 & 0.341331 & -1.17092 & 3.83355 \\
HF &-1.13031 (0.02010)& 1.38214 (-0.04590)& 0.410051 (0.06872)&  -1.35488 (-0.18396)& 4.56773 (0.73418)\\
$E_K^{2B}[G_{HF}]$ &-1.15070 (-0.00029) &  1.43037 (0.00233) & 0.336672 (0.00466)& -1.16647 (0.00445) & 3.79075 (-0.04280)\\
$E_K^{2B}[G_{\textrm{mod}}]$ &-1.14960 (0.00081) & 1.42190 (-0.00615) & 0.353482 (0.01215) & -1.18505 (-0.01413) &  3.93920 (0.10564) \\
$E_K^{GW}[G_{HF}]$ &  -1.15883 (-0.00842) & 1.42830 (0.00025) & 0.348166 (0.00683) & -1.16481 (0.00612) & 3.83302 (-0.00053) \\
$E_K^{GW}[G_{\textrm{mod}}]$ &
-1.15772 (-0.00731)& 1.42145 (-0.00659) & 0.360504 (0.01917) & -1.18546 (-0.01454)& 3.93862 (0.10507) \\
$E_K^{T}[G_{HF}]$ & -1.14689 (0.00352) & 1.41277  (-0.01528)& 0.365893 (0.02456) & -1.22590 (-0.05498)& 4.06337 (0.22981) \\
$E_K^{T}[G_{\textrm{mod}}]$ &  -1.14617 (0.00424) & 1.40817 (-0.01987) & 0.374923 (0.03359)& -1.23936  (0.06844) & 4.14125 (0.30770) \\
$E_{LW}^{2B}[G_{HF}]$ & -1.14955 (0.00086) & 1.42133 (-0.00672) & 0.355003 (0.01367)& -1.18399 (-0.01307)& 3.95963 (0.12608)\\
$E_{LW}^{2B}[G_{\textrm{mod}}]$ & -1.14964 (0.00077)&1.42239 (-0.00566)& 0.352213 (0.01088)& -1.18552 (-0.01460)& 3.92456 (0.09100) \\
$E_{LW}^{GW}[G_{HF}]$ & -1.15747 (-0.00706) & 1.42110 (-0.00694) & 0.360164 (0.01883) & -1.18946 (-0.01854)& 3.93999 (0.10643) \\
$E_{LW}^{GW}[G_{\textrm{mod}}]$ & -1.15754 (-0.00713)&1.42173 (-0.00631)& 0.358821 (0.01749) & -1.18820 (-0.01728) & 3.92913  (0.09558)\\
$E_{LW}^{T}[G_{HF}]$ & -1.14633 (0.00408)& 1.40943 (-0.01861) & 0.372198 (0.03087)& -1.23644 (0.06552)& 4.11844 (0.28488) \\
$E_{LW}^{T}[G_{\textrm{mod}}]$ & -1.14636 (0.00405) & 1.40968 (-0.01836) & 0.371611 (0.03028) & -1.23610 (-0.06518) & 4.11337  (0.27982)\\
SC-2B & -1.14241 (0.00799) & 1.40065 (-0.02739) &  0.379490 (0.03815)&& \\
SC-GW & -1.14940 (0.00100)& 1.40480 (-0.02324) & 0.376746  (0.03541)&& \\
SC-T & -1.13830 (0.01210)&  1.39114 (-0.03690)& 0.395308 (0.05398) &&\\
\hline
\end{tabular}
\end{ruledtabular}
\end{table}


\section{LW correction for the model input}
\label{LW_model}
\begin{figure}[ht]
  \subfloat[LW correction for the 2B approximation with the HF and the model input Green's function (in red and blue respectively)]{\includegraphics[width=0.45\textwidth]{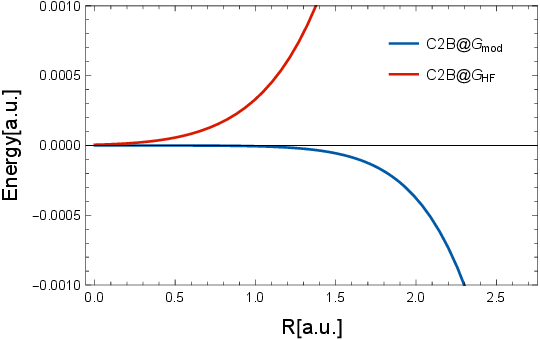}\label{fig:f1correction}}
  \hfill
  \subfloat[LW correction for the T-matrix approximation with the HF and the model input Green's function (respectively in red and blue respectively) ]{\includegraphics[width=0.45\textwidth]{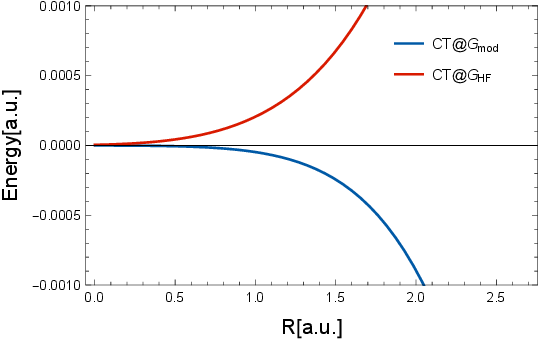}\label{fig:f2correction}}
  \caption{The LW corrections for the 2B and T-matrix approximation, where it can be seen that $C_\LW [G_{\textrm{mod}}] <0$ for both the 2B and the T-matrix}
    \label{LW_2BT_correction}
\end{figure}

\begin{figure}[htp]
\includegraphics[width=0.45\textwidth]{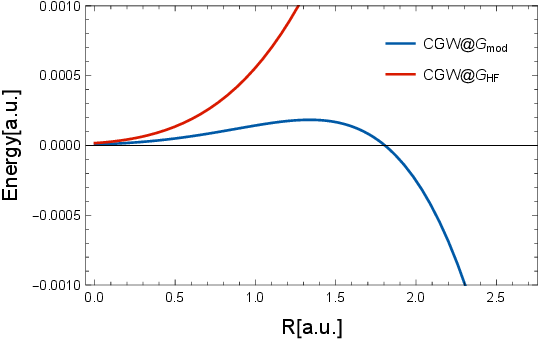} 
\caption{LW correction for GW with the HF and the model input (red and blue curves respectively). We see that $C_\LW^\GW [G_{\textrm{mod}}]$ can become positive for small values of $R$}
\label{LW_GW_correction}
\end{figure}

In this section we explicitly show that the sign of the LW correction used in combination with the model Green's function is negative for the 2B and T-matrix approximations. 
Factoring out a common denominator in 
Eq.(\ref{eq:tG_correction}) one finds:
\begin{align}
C_\LW^\mx [G_{\textrm{mod}}] &= \frac{4\gamma^\mx B^\mx}{(\Delta + \Omega^\mx)(\delta + \Omega^\mx + \sqrt{(\delta+ \Omega^\mx)^2 + 4 \gamma^\mx})},
\end{align}
with $B^\mx$ defined by:
\begin{align}
B^\mx &= \delta + \Omega^\mx + \sqrt{(\delta+ \Omega^\mx)^2 + 4 \gamma^\mx} - 2 (\Delta + \Omega^\mx ) \nonumber \\
&=  4t -2 \delta + \sqrt{(\delta+ \Omega^\mx)^2 + 4 \gamma^\mx} -(\delta + \Omega^\mx)
\end{align}
where we used that $\Delta =2 \delta - 2t$.
Now, since $\delta \geq  2t$ and $\Omega^\mx \geq \Delta \geq 2t$ such that $\delta + \Omega^\mx \geq 4t$ and it follows that:
\be
\sqrt{(\delta+ \Omega^\mx)^2 + 4 \gamma^\mx} -(\delta + \Omega^\mx) \leq \sqrt{(4t)^2 + 4 \gamma^\mx} -4t
\ee
since the function $\sqrt{x^2+4 \gamma^\mx}-x$ is a monotonically decreasing function of $x$.
Equipped with this inequality we find:
\begin{align}
B^\mx  \leq -2 \delta &+ 2 \sqrt{(2t)^2 +  \gamma^\mx}  \\
&= 2 [ \sqrt{(2t)^2 +  \gamma^\mx} - \sqrt{ (2t)^2  +(U-w)^2/4} ] \nonumber
\end{align}
From the above expression it follows that $B^\mx \leq 0$ and therefore $C_\LW^\mx [G_{\textrm{mod}}] \leq 0$ whenever $\gamma^\mx \leq (U-w)^2/4$. An inspection of the specific form of $\gamma^\mx$ for our approximations reveals that this condition is exactly satisfied for the 2B and T-matrix approximations.
For the GW case, we only have $\gamma^\GW \leq (U-w)^2/2$ and calculations show that in this
case, $C_\LW^\mx [G_{\textrm{mod}}]$ can become positive, as is displayed in Fig.\ref{LW_GW_correction}. \\
Instead Fig.\ref{LW_2BT_correction} clearly shows that for 2B and T-matrix $C_\LW^\mx [G_{\textrm{mod}}]$ is negative. The underlying reason is that both 2B and T-matrix correctly include the second order exchange diagram which makes the condition $\gamma^\mx \leq (U-w)^2/4$ valid, whereas such diagrams is absent in GW, which thereby overestimates the value of $\gamma^\mx$.

 \clearpage

\bibliography{biblio-lw}

\end{document}